\documentclass[12pt,preprint]{aastex}

\newcommand{\bdv}[1]{\mbox{\boldmath$#1$}}

\def\au{{\rm AU}} 
\def\sinc{{\rm sinc}} 
\def\kms{{\rm km}\,{\rm s}^{-1}}
\def\masyr{{\rm mas}\,{\rm yr}^{-1}}
\def\kpc{{\rm kpc}}
\def\var{{\rm var}}

\def\orb{{\rm orb}}
\def\obs{{\rm obs}}

\def\rel{{\rm rel}}
\def\ast{{\rm ast}}

\def\balpha{{\bdv\alpha}}

\def\bp{{\bf p}}
\def\hatn{{\bf \hat n}}
\def\erf{{\rm erf}}
\def\rot{{\rm rot}}
\begin{document}
\title{Microlens Surveys are a Powerful Probe of Asteroids}

\author{Andrew Gould \& Jennifer C.\ Yee}
\affil{Department of Astronomy, Ohio State University,
140 W.\ 18th Ave., Columbus, OH 43210, USA; 
gould,jyee@astronomy.ohio-state.edu}

\begin{abstract}

While of order a million asteroids have been discovered, the number
in rigorously controlled samples that have precise orbits and rotation
periods, as well as well-measured colors, is relatively small.
In particular, less than a dozen main-belt asteroids with estimated diameters
$D<3\,$km, have excellent rotation periods.
We show how existing and soon-to-be-acquired
microlensing data can yield a large asteroid
sample with precise orbits and rotation periods, which will include roughly
6\% of all asteroids with maximum brightness $I<18.1$ and lying within
$10^\circ$ of the ecliptic.  This sample will be dominated by small
and very small asteroids, down to $D\sim 1\,$km.  We also show how
asteroid astrometry could turn current narrow-angle OGLE proper motions
of bulge stars into wide-angle proper motions.  This would enable
one to measure the proper-motion gradient across the Galactic bar.

\end{abstract}

\keywords{gravitational lensing: micro --- minor planets, asteroids} 

\section{{Introduction}
\label{sec:intro}}

Modern surveys are discovering asteroids in prodigious numbers, soon
to exceed a 
million\footnote{http://www.minorplanetcenter.net/iau/MPCORB.html}.
The applications of such discoveries
range from study of the detailed structure of Solar System resonances to
defense from (or at least evacuation in the face of) asteroid
collisions with Earth.  However, despite the exponential increase
in asteroid detection, the number of asteroids with well-measured rotation
periods and colors (as well as as orbital elements) remains quite small.
\citet{warner09} have assembled a catalog of minor planets with at
least some rotation-period information,
which in its most recent (November 2012) edition has 5877 
entries.\footnote{http://www.minorplanet.info/lightcurvedatabase.html}
Among these are 1113 main-belt asteroids with excellent period
determinations (quality flag `3'; excluding `3-').

\begin{figure}
\plotone{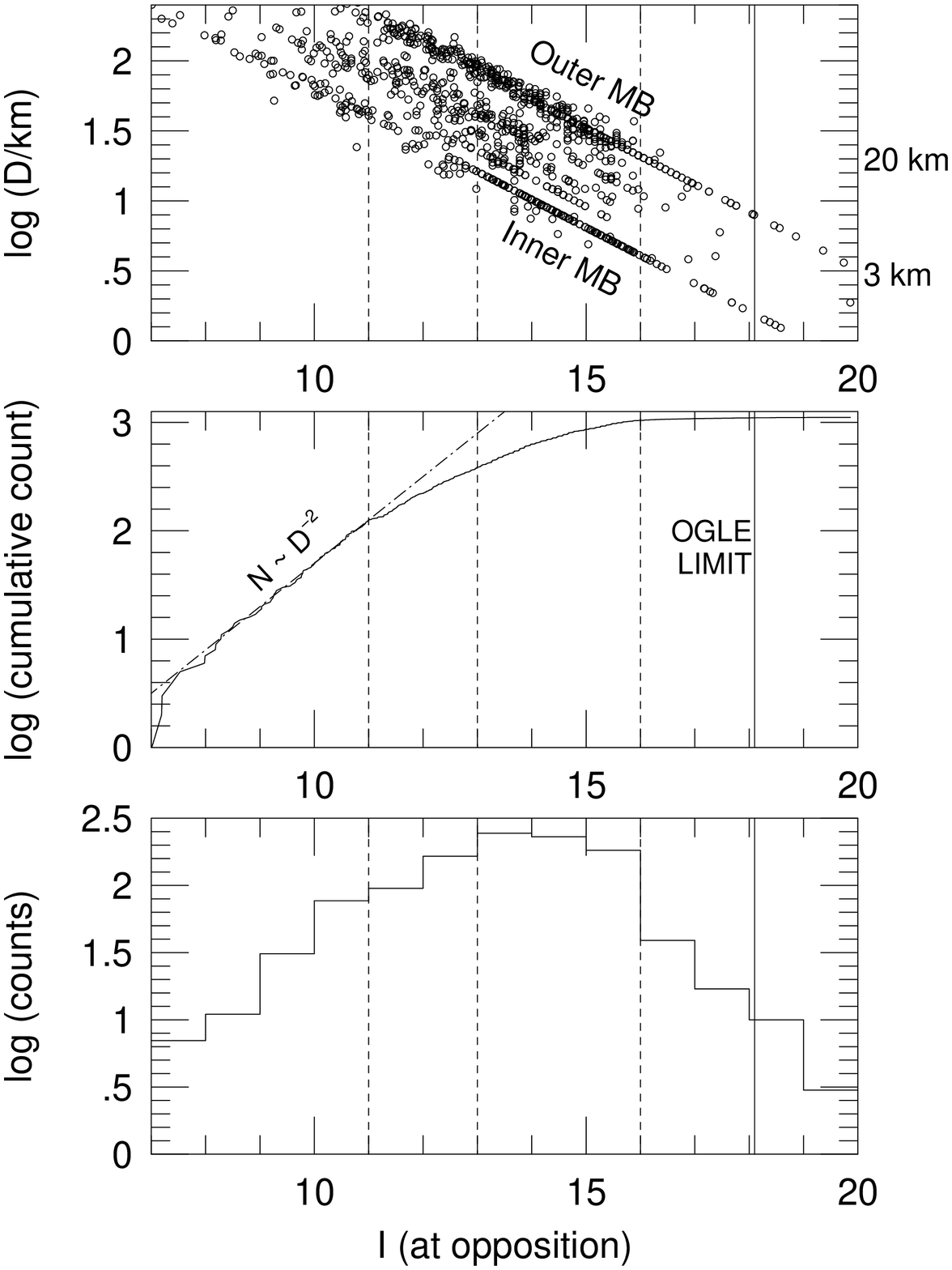}
\caption{\label{fig:cumcnt}
Lower and middle panels: Differential and cumulative number counts 
of asteroids with excellent rotation-period 
determinations as a function of estimated $I$-band magnitude
at opposition as derived from the Nov 2012 update of
\citet{warner09}.  Apparent breaks in the distribution are indicated
by dashed vertical lines: below $I=11$, the distribution obeys a
diameter distribution $N\propto D^{-2}$ (dot-dashed line), it begins
to turn over at $I=13$, and suffers a rapid
drop-off for $I>16$.  This is compared to our estimated completeness
limit from an OGLE-based survey of $I=18.1$.  Upper panel: estimated
diameters of these same cataloged asteroids.  Very few are smaller
than 3 km.
}
\end{figure}

Figure \ref{fig:cumcnt} shows three different views of this sample.
The lower and middle panels show differential and cumulative counts as a 
function of estimated $I$-band magnitude at opposition.  For this purpose,
we assumed that inner, middle, and outer main-belt asteroids are
at distances of 2.3, 2.9, and 3.4 AU from the Sun.  We also assumed
$(V-I)=0.7$ (similar to the Sun).  That is, 
$I= H + 5\log(x(x-1))-0.7$, where $H$ is the the cataloged absolute
magnitude and $x\au$ is the assumed heliocentric distance.
There is a clear break at about $I=11$, a rounding commencing at about
$I=13$, and a sharp drop-off for $I>16$.
The upper panel shows that this drop-off corresponds to asteroid
diameters of $D\sim 3\,$km in the inner main belt and about $D\sim 20\,$km
in the outer main belt.  Hence, while orbital elements are being measured
for vast numbers of asteroids, the rotation periods of small main-belt asteroids
are not being probed by current studies.

Observations of Near Earth
Asteroids show that objects down to diameters of $D\sim200$m have
a minimum rotation period of 2.2 hrs \citep{Pravec07}, presumably
because a faster spin exceeds the gravitational binding force. At
present there are insufficient numbers of main belt asteroids with
measured rotation periods and $D<3$ km to determine whether they also obey 
this spin barrier. \citet{Dermawan04} attempted to address this question
by measuring the rotation periods of small main-belt asteroids.  However
\citet{warner09} concluded that none of the rotation periods are secure,
i.e., have a quality factor `3'.

The fundamental reason that only a tiny fraction ($\sim 10^{-5}$) of the
small asteroids being discovered have good rotation periods is that
the discovery photometry is very sparse and followup photometry is rare.  
By contrast, modern bulge microlensing
surveys have quite dense sampling, often 10--25 epochs per night.
Moreover, these high-cadence zones cover contiguous areas that span
$4^\circ$ or more, meaning that typical asteroids that intersect these
fields spend several weeks within them.

One fairly simple idea then, would be just to identify known asteroids
passing through these fields and measure their rotation periods (and colors)
using archival microlensing data.  However, microlensing surveys can
have considerably greater return than this for asteroid science.  We show
here that these surveys can, by themselves, completely characterize the
orbits of a large, well-defined subset of asteroids $I<18.1$ 
that pass through their high-cadence fields.
This means that it is possible to assemble a sample of asteroids that
have not only well-defined orbits, rotation periods and colors, but also
well-defined selection.  By contrast, the present sample of 
small-to-moderate size asteroids with both rotation periods and orbits is 
completely heterogeneous.

Thus, present and future microlensing surveys could probe completely
new regions of asteroid parameter space.

\section{{Asteroids in Microlensing Surveys}
\label{sec:asteroids}}

Our focus here is the possibility of extracting a large statistically
well-defined sample of asteroids from existing and planned microlensing 
data.  By ``statistically well-defined'', we mean primarily a sample
whose orbital elements and selection procedure are well enough
understood that the sample can be rigorously compared
with a population model that specifies the distributions of orbital
elements and (solar system) absolute magnitudes.  The (microlensing)
data are taken completely without reference to the possibility that
asteroids will be or have been detected.  Therefore, the key question
is: how well can the orbital parameters be measured from existing data?
Hence,
we will focus mainly on the question of whether orbital elements can
be accurately measured from microlensing data alone for a substantial
number of asteroids.

\subsection{{Framing the problem}
\label{sec:frame}}

Microlensing surveys target the Galactic bulge, with the center
of high-cadence observations being near
(RA,Dec) = (17:58:00,$-29$:20:00),
i.e., $6^\circ$ south of the Winter Solstice.  The size of the high-cadence
region varies according to survey.   In this study we will adopt
the parameters of the OGLE-IV survey and will comment in 
Section~\ref{sec:other} on how the
results should be adjusted for other surveys.  OGLE-IV surveys an area
of 11 contiguous square degrees at least once per hour, with 3/8 of this
area being observed three times per hour.  Typical trajectories crossing
this zone would intersect it for about $4^\circ$.  This high-cadence
region is embedded in a larger low-cadence 
area\footnote{http://ogle.astrouw.edu.pl/sky/ogle4-BLG/}, with one or more
observations per night (May--August) over a contiguous area
of about 50 deg$^2$.
These observations are carried out whenever the
bulge is visible, except three nights per month, typically chosen to
be those when the Moon is passing through the bulge.  Seeing of $1''$
or somewhat higher is achieved on more than half of all nights.  

\begin{figure}
\plotone{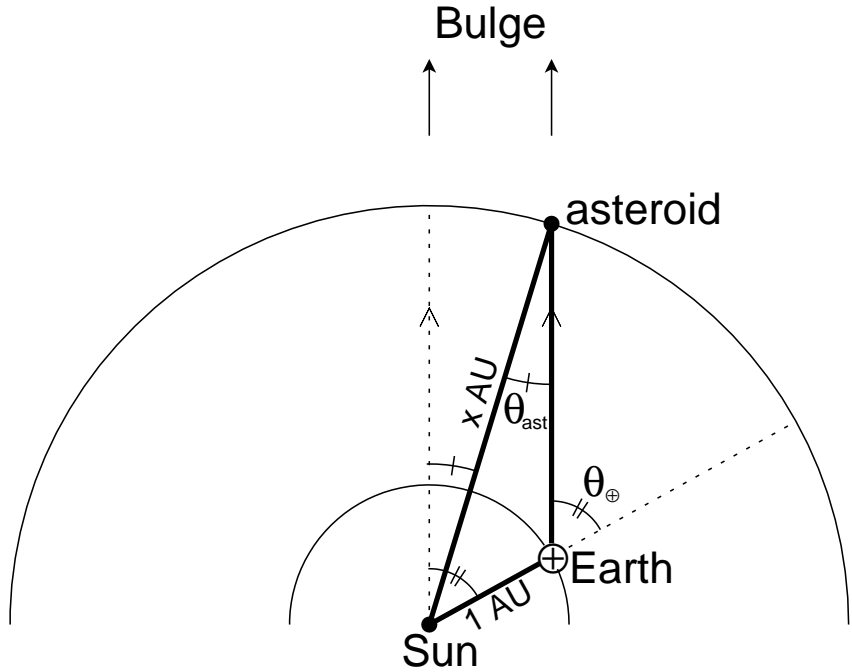}
\caption{\label{fig:geom}
Idealized geometry of asteroids passing through bulge microlensing fields.
As the Earth moves (orbital phase $\theta_\oplus$) in its approximately
circular orbit, the observations target the same bulge fields.  The
angle bulge-Sun-asteroid to an asteroid at $x\au$ can
therefore be calculated from the law of sines: 
$\sin\theta_{\rm ast} = x^{-1}\sin\theta_\oplus$.  If the asteroid orbit
is approximated as circular, then at 
$\theta_\oplus = \theta_{\oplus,{\rm stat}} = \sin^{-1}[x(1+x+x^2)^{-1/2}]$
the asteroids transition from prograde to retrograde motion (or vice versa).
See Equation~(\ref{eqn:stationary}).  Specifically,
$49^\circ<\theta_{\oplus,\rm stat}<61^\circ$ for $2<x<4$.  The season when many new
asteroids can be discovered is approximately $2\theta_{\oplus,{\rm stat}}$.
The fraction of asteroids discovered during this time is
$f_{\rm season}= [\theta_{\rm ast,stat}-\theta_{\oplus,\rm stat}x^{-3/2}]/\pi$, i.e.,
2.7\%--2.9\% per season.
}
\end{figure}

Figure~\ref{fig:geom} shows the geometry of highly idealized asteroids
in co-planar circular orbits.  The angles bulge-Sun-Earth 
and bulge-Sun-asteroid are labeled $\theta_\oplus$ and $\theta_\ast$,
respectively.  By the law of sines
\begin{equation}
\sin\theta_\oplus = x\sin\theta_\ast
\label{eqn:lawsines}
\end{equation}
where $x\au$ is the radius of the asteroid orbit.  The transverse velocity
of the asteroid relative to the Earth is then
\begin{equation}
v_\rel= (v_\ast - v_\oplus)_\perp = v_\ast\cos\theta_\ast - v_\oplus\cos\theta_\oplus .
\label{eqn:vrel}
\end{equation}
Setting $v_\rel$ to zero in this equation and combining it with 
Equation~(\ref{eqn:lawsines}) and Kepler's Third Law
yields the orbital phases at which the asteroids are stationary (the transitions
from prograde-to-retrograde and retrograde-to-prograde motion)
\begin{equation}
\sin\theta_{\oplus,\rm stat}=x\sin\theta_{\ast,\rm stat}=\pm\sqrt{x^2\over 1 + x + x^2}.
\label{eqn:stationary}
\end{equation}
Note that for $x=(2,3,4)$, we have 
$\theta_{\oplus,\rm stat}= \pm(49^\circ,56^\circ,61^\circ)$.  Thus, to a good
approximation, this defines a zone of retrograde
motion during $\pm 60$ days of the Summer Solstice (21 June).
We now argue that
to a good approximation, asteroids will be discovered
if and only if they pass through the high-cadence fields 
during this ``retrograde season''.

There are a number of factors that vary over the course of the
``retrograde season''. We will begin by discussing observability and its
affects on photometric and astrometric measurements. 
In Sections~\ref{sec:determ} and \ref{sec:variations}
we will examine the impact of additional effects, such as asteroid
phase angle and degree of trailing.
First note that at the
Summer Solstice (and for 15 days on either side) the bulge fields can
be observed at airmass $<2$ for 10 hours, centered on their transit of
the zenith.  At 60 days from Summer Solstice, they can be observed for
about 4.5 hours on one side of transit and 2 hours on the other.
Observations near Summer Solstice therefore have a modest advantage of
a longer observing window, which we will see in
Section~\ref{sec:single} scales as $t_\obs^{3/2}$, i.e., a factor
$(10/6.5)^{3/2} = 1.9$.  However, near their turning point the
asteroids are moving more slowly and therefore spend more time in the
microlensing fields.  This effect scales as $v_\rel^{-3/2}$. Of
course, real asteroids are not on circular co-planar orbits.  They
could not be observed $6^\circ$ from the ecliptic if they were.  But
the point is these two factors go in opposite directions.

On the other hand, most of the asteroids that ``enter'' the high-cadence
fields just after the ``retrograde season'' has ended are in fact
re-entering and were already discovered during the retrograde season.
Similarly, those that exit just before the start of the retrograde season,
will enter the high-cadence fields at a later point, during the
retrograde season.  So there are few additional discoveries during these
periods.  And further from the edges of the retrograde season, 
the bulge observation window is rapidly contracting,
making orbit characterization difficult.  Of course, some very interesting
asteroids will be discovered during this period, including asteroids 
interior to the Earth's orbit and other difficult-to-detect asteroids.
However, from the point of view of characterizing the main features of
the survey, these are a distraction.  

Hence, since the great majority of discoveries will take place during
$\pm 60\,$days of opposition (Summer Solstice), and since survey geometry
does not vary much over this period, we will initially carry out
our calculations at Summer Solstice, when the asteroids are at opposition.

Finally we note that while Earth is traveling from 
$-\theta_{\ast,\rm stat}$ to $+\theta_{\ast,\rm stat}$, the bulge-field observations
cover an arc $2\theta_{\ast,\rm stat}$ of an asteroid orbit.  At the same
time, the asteroids move (in the same direction)
through an arc $2\theta_{\oplus,\rm stat} x^{-3/2}$.
Therefore, one season of observations covers a fraction of all 
asteroids in these orbits of
\begin{equation}
f_{\rm season} = {\theta_{\ast,\rm stat} - \theta_{\oplus,\rm stat} x^{-3/2}\over \pi}.
\label{eqn:fseason}
\end{equation}
For $x=(2,3,4)$, we have $f_{\rm season}=(0.027,0.029,0.028)$, i.e., roughly
constant.  Since, the ``cross section'' of the high-cadence fields is
about $4^\circ$, this means that a 10-year survey of this type would
detect roughly 
$(0.028\,{\rm yr}^{-1})(10\,{\rm yr})(4^\circ/20^\circ)\sim 6\%$ 
of all asteroids passing
within $\pm 10^\circ$ of the ecliptic that satisfy the magnitude limit
(still to be established).

Virtually all asteroids discovered in this fashion would have
$(V-I)$ colors from OGLE data, and lightcurves spanning of order
40 days (i.e., not just in the high-cadence regions).  Since
asteroid rotation periods are typically less than one day, these
observations would be sufficient to measure the periods, as well
as amplitudes of variation, with good precision.
Hence they can also be used to identify binaries and measure asteroid
shapes.

\subsection{{Asteroid Detection}
\label{sec:detect}}

Microlensing surveys use difference imaging
\citep{alard98,wozniak00}, in which a reference image is
convolved to the seeing of each target image, and then subtracted from it,
which removes all non-varying (constant) sources.  All that remains are 
``difference stars'' at locations where
the flux has changed.  These include microlensing events and other
stars whose brightness has changed, but it also includes asteroids,
whose positions have changed.  Microlensing fields are quite crowded,
which means that the asteroid will frequently overlap with field
stars.  However, the great majority of these stars are below the sky
background
and so will subtract out in the difference image without even adding
appreciable noise above the general background.  Hence, the signature
of an asteroid in these microlensing fields 
is the same as it would be for a series of images
taken of a high-latitude field, i.e., a moving object.  The only difference
is that bulge fields are more crowded, so occasionally the signal is degraded
by confusion with bright or variable stars.

In principle, one could think about finding asteroids that are
too faint to detect in individual images by stacking images along
(hypothetical) asteroid orbits.
However, as we will show below, it is actually impossible to properly
characterize the orbits of asteroids that are undetectable in individual
images (using microlensing data alone).  
Since ``typical'' asteroids are moving at retrograde
velocities $v\sim 12\,\kms$ at distances $r\sim 2\,\au$ from Earth, they
move only $\sim 30''$ per hour.  Hence, in fields with observations at 
least once per hour, it is straightforward to identify
such tracks on individual nights.  Associating tracks between nights
is also not difficult.  As we will show in Section~\ref{sec:pred}, 
from a single
night of data one can predict the position on the next night to
within ${\cal O}(3'')$, which is extremely precise compared to the 
surface density of asteroids.

\section{{Orbit Determination}
\label{sec:determ}}

\subsection{{Orbit Parameterization}
\label{sec:params}}

Defining an orbit usually means measuring the six Kepler parameters,
i.e., the five invariants of motion plus the orbital phase.  However,
from the standpoint of understanding the orbital errors, it is more
convenient to study the mathematically equivalent parameterization of
six Cartesian phase-space coordinates at a given instant of time,
i.e., three positions and three velocities.  Since the Earth's orbit
is known extremely precisely, these six coordinates are equivalent
to the following six: 2 instantaneous angular positions, 2 instantaneous
angular velocities, the instantaneous distance, and
the instantaneous ``radial velocity'', all measured from the center of the
Earth.  If these six are
specified, one can transform to Kepler parameters, and vice versa.

Hence, suppose that one conducts observations over 20 days while an
asteroid passes through the high-cadence fields and then considers the
ensemble of Kepler models that are consistent with these data. In
principle, these models might span a large range of Kepler parameters,
but all will agree extremely precisely on the angular positions as
seen from the center of the Earth at the instant that the field
transits on each night of observations.  This is because these
quantities are very nearly direct observables.  That is, any model
that is consistent with the data
must reproduce the arc of mean nightly positions of the asteroid, and
hence will yield the same angular positions and velocities at a given
instant.  

These precise nightly determinations have two distinct implications.
First they imply that the angular positions and velocities
at some fiducial instant near the midpoint of observations are extremely
well determined (four parameters).  Second, they enable a particularly
simple estimate of the measurement precision of the remaining two
parameters, i.e., the distance and radial velocity at this same instant.
This is because, conceptually, fixing the geocentric angular position 
and velocity enables one to determine the diurnal parallax offset from 
each individual
astrometric measurement, rather than carrying out a fit to the entire
data set.  The asteroid-distance measurement on a single night is then
a simple integral over these measurements.  The standard error of the
mean of these measurements then gives the error in the distance at the above
fiducial instant, while the error in the slope is the radial-velocity
error.  Hence we must begin by deriving the distance error from a single
night of data {\it given} that we know the angular position and velocity
at transit.

\subsection{{Single-night distance measurements}
\label{sec:single}}

The angular displacement of the asteroid due to diurnal motion
of the observatory is $\Delta\balpha=\Delta \bp/r$ where
$\bp = (p_n,p_e)$ is the projected position of the observatory relative to
the center of the Earth and $r$ is the asteroid distance.  While
$\bp$ changes in both the north and east directions during the night,
the amplitude of the changes in the north component
is generally small and will be ignored
here in the interest of simplicity.  Hence,
the precision of the parallax measurement $\pi\equiv \au/r$ 
to an asteroid at distance $r$ from Earth
derived from $n$ measurements of astrometric precision
$\sigma_0$ is  given by
\begin{equation}
\sigma(\pi) = \pi{\sigma(r)\over r}=
{\au\sigma_0\over \sqrt{n\,\var(p_e)}} = 
2.2\times 10^{-3}\biggl({n\over 10}\biggr)^{-1/2}
\biggl({\zeta_e\over 0.5}\biggr)^{-1}
\biggl({\sigma_0\over 30\,\rm mas}\biggr)
\label{eqn:parallaxerr}
\end{equation}
where $\zeta_e\equiv [\var(p_e)]^{1/2}/R_\oplus$ is the standard deviation
of observatory motion relative to the Earth center in the east direction
normalized to $R_\oplus$.  Hence, for example, the distance of
an asteroid at $r=2\,\au$ would be measured to a precision of 
$8.8\times 10^{-3}\au$, assuming the above fiducial parameters.
Note that for observations spanning $t_\obs$ and centered on transit
\begin{equation}
\zeta_e \sim \sqrt{1 -\sinc(2\pi t_\obs/{\rm day})\over 2}\cos\delta
\rightarrow 0.55
\label{eqn:etae}
\end{equation}
where $\delta$ is the latitude of the observatory and 
where we have assumed
that the asteroid declination is similar.  The final evaluation
assumes $t_\obs=10\,$hr and $\delta=-30^\circ$.

Note that in the regime of interest, 
$1-\sinc(y)\sim y^2/6\propto t_\obs^2$.  Since $n\propto t_\obs$,
this implies $\sigma(\pi)\propto t_\obs^{-3/2}$ as claimed above.
More detailed calculations show that this relation remains approximately 
valid for a time series whose center is somewhat offset from transit,
such as those at the limits of the retrograde season.

\subsection{{Radial velocity measurement}
\label{sec:radial}}

A typical asteroid at opposition moves at about $0.2^\circ$ per day.
Hence it remains in a $4^\circ$ field for about $T_\obs=20\,$days.
Assuming a fraction $f_{\rm gd}$ of the nights have good seeing
and low background, and that these $m=f_{\rm gd}T_\obs/{\rm day}$
nights are roughly uniformly distributed over $T_\obs$, then the
mean radial velocity will be measured with precision
\begin{equation}
\sigma(v_r) = \sqrt{12\over m}{\sigma(\pi)r^2\over T_\obs\au}
=0.81\,\kms\biggl({n\over 10}\biggr)^{-1/2}
\biggl({\zeta_e\over 0.5}\biggr)^{-1}
\biggl({\sigma_0\over 30\,\rm mas}\biggr)
\biggl({f_{\rm gd}\over 0.5}\biggr)^{-1/2}
\biggl({T_\obs\over 20\,{\rm days}}\biggr)^{-3/2}
\biggl({r\over 2\,\au}\biggr)^{2}.
\label{eqn:sigmavr}
\end{equation}
By contrast, the error in the mean distance (over the measured
trajectory) will be $\sigma(\bar r)= m^{-1/2}\sigma(\pi)r^2/\au$.  
Hence, the ratio of these two errors will be
\begin{equation}
{\sigma(v_r)/v_\oplus\over \sigma(\bar{r})/\au}
=\sqrt{12}{\au\over v_\oplus T_\obs} = 9.9
\biggl({T_\obs\over 20\,{\rm days}}\biggr)^{-1}.
\label{eqn:errrat}
\end{equation}
Thus, for typical asteroids at $a\sim 3\au$, 
$\sigma(v_r)/v_\ast \sim 17 \sigma(\bar r)/a$.  This implies that the
radial-velocity error is by far the dominant error in the problem.
Relative to this error, the five other phase-space coordinates are
known with essentially infinite precision.

\subsection{{Impact on orbital period estimate}
\label{sec:period}}

In the above approximation that all instantaneous Cartesian parameters 
except $v_r$ are known perfectly, the specific potential energy is also known
perfectly, while the fractional uncertainty in the specific kinetic energy
is $2 v_r\sigma(v_r)/v_{\rm \ast}^2$, where now $v_r$ is taken to mean
the asteroid velocity in the direction of the Earth but the frame
of the Sun.  Hence, the fractional error in the orbital period $P$ is
\begin{equation}
{\sigma(P)\over P}={3\over2}{\sigma (a)\over a}=
{3\over2}{\sigma (|E|)\over |E|}\sim 3{v_r\sigma(v_r)\over v_\ast^2}
= 0.028{\sigma(v_r)\over 0.81\,\kms}\,{v_r/v_\ast\over 0.2}
\biggl({v_\ast\over 17\,\kms}\biggr)^{-1}.
\label{eqn:perioderror}
\end{equation}
Thus, for the above fiducial parameters and a $P=5$ year orbit, 
$\sigma(P)\sim 50\,$days.

\subsection{{Astrometric precision and systematic errors}
\label{sec:astrometric}}

The above estimates were scaled to astrometric measurements of 30 mas.
This is a typical error for good-seeing, low-background OGLE-IV 
stellar images at
$I\sim 18.4$ (R.\ Poleski, private communication 2012).
However, asteroid images are not stellar, but trailed.  We
evaluate the impact of trailing in the Appendix in terms of
$\eta\equiv \epsilon/2\sigma$, where $\epsilon$ is the length of
the trail and $\sigma$ is the Gaussian width of the point spread
function (PSF).  This impact is more severe for astrometric
measurements in the direction of trailing than the orthogonal
direction.  Because the astrometry is limited by diurnal parallax
measurements, which are essentially East-West and therefore
aligned with the direction of asteroid motion (and so trailing),
we evaluate the astrometric degradation factor $G_x^{1/2}(\eta)$
in the trailing direction.  For our fiducial geometry, i.e.,
observations at opposition of an asteroid at $x=3$ and reasonably
good seeing $\sigma=0.5^{\prime\prime}$ and 100 s OGLE exposures,
we find $\eta=1$ and therefore $G_x^{1/2}(\eta)=0.8$.  Hence our
fiducial precision is achieved at $I=18.1$.

One must also worry about systematic errors, the most
important of which is differential refraction.  Asteroids are 1--3 mag
bluer in $(V-I)$ than the red giants that mainly define the astrometric
reference frame in bulge fields, and the scale of the 
differential refraction is 
$\sim 7\,$mas per mag per airmass.  Differential refraction 
is extremely important because
the distance measurements in Section~\ref{sec:single} were based on
measuring the astrometric deflection as a function of position
relative to the Earth's center.  The latter is very strongly correlated
with airmass.  Now, differential refraction can be calibrated from
stars of similar color to the asteroids.  
However, (R.\ Poleski, private communication 2012)
finds a systematics error floor of $\sigma_{\rm sys} \sim$ 2--3 mas, 
even after correction.

Since these errors are systematic, one cannot necessarily rely on reducing them
according to ``square-root of $N$'', as was assumed in the equations
for $\sigma(\bar r)$ and $\sigma(v_r)$.  However, the error floor imposed
by systematic errors is very different in the two cases.  The measurement
of $\sigma(\bar r)$ will not further improve if 
$\sigma_0<\sqrt{nm}\sigma_{\rm sys}$, whereas for $\sigma(v_r)$ the
improvement stops for $\sigma_0<\sqrt{n}\sigma_{\rm sys}$.  Since
the orbit determination is fundamentally limited by the latter,
this implies that statistical precision is limited at 
$\sigma_0 = \sqrt{n}\sigma_{\rm sys}\sim 10\,$mas.  Since this
is substantially below the fiducial value adopted in 
Equation~(\ref{eqn:parallaxerr}), the impact of systematic errors
is likely to be small.

\subsection{{Precision of rotation periods and amplitudes}
\label{sec:rotpers}}

Just as the astrometric precision is degraded by trailing, so is
the photometric precision.  However, this effect
has the same magnitude as the
astrometric degradation in the orthogonal direction, $G_y^{1/2}(1) = 0.93$.

At $I=18.4$ typical OGLE photometry errors are $\sigma(I)\la 0.05\,$mag.
therefore, we adopt $\sigma(I)\la 0.04\,$ mag at $I=18.1$ including the 
effect of trailing.  Small asteroids that would be observed near this limit 
have (full) amplitudes $A$ of 0.1--0.2 mag.  To evaluate the period and
amplitude errors, we adopt a simple sinusoidal model:
$I(t)=I_0[1+(A/2)\cos(2\pi t/P_\rot + \phi)]$. For $N$, roughly uniformly 
sampled data points over time $T$, the amplitude error is then
$[\sigma(A)]^{-2} = (N/T)\int dt (d I/dA)^2$ or
\begin{equation}
{\sigma(A)\over A} = 0.075
{\sigma(I)\over 0.04}
\biggl({A\over 0.15}\biggr)^{-1}
\biggl({n\over 10}\biggr)^{-1/2}
\biggl({f_{\rm gd}\over 0.5}\biggr)^{-1/2}
\biggl({T_\obs\over 20\,{\rm days}}\biggr)^{-1/2},
\label{eqn:amp}
\end{equation}
which will therefore typically be quite good.  

Similarly, the fractional error (or rather, strictly speaking,
the minimum variance bound) in the rotation period is given by
$[\sigma(P_\rot)]^{-2} = (N/T)\int dt (d I/d P_\rot)^2$ or
\begin{equation}
{\sigma(P_{\rm rot})\over P_{\rm rot}} = 0.002
{P_{\rm rot}\over 1\,{\rm day}}
{\sigma(I)\over 0.04}
\biggl({A\over 0.15}\biggr)^{-1}
\biggl({n\over 10}\biggr)^{-1/2}
\biggl({f_{\rm gd}\over 0.5}\biggr)^{-1/2}
\biggl({T_\obs\over 20\,{\rm days}}\biggr)^{-3/2}
\label{eqn:per}
\end{equation}

However, while the amplitude can be estimated directly from the scatter
(assuming that the measurement errors are known), the rotation period
must be found by testing all possible folds of the lightcurve, via e.g.,
a periodogram or Fourier transform.  This can lead to multiple minima,
and hence unless a single period is decisively favored by these tests,
the true uncertainty can be much larger than the minimum variance bound.
Thus, real data must be fitted before the true period completeness can
be assessed as a function of $P_\rot$, $A$, and $\sigma(I)$.
See for example \citet{warner11,harris12}.

\section{{Variation of Sensitivity}
\label{sec:variations}}

For simplicity, we have so far evaluated the sensitivity of microlensing
surveys under the highly idealized assumption that detections are
made at opposition and that asteroids are typically at distance $r=2\,\au$
($x=3$).  We now successively relax these two assumptions.

As the ``retrograde season'' progresses from opposition toward prograde
motion, there are four distinct  changes that affect the survey
sensitivity: 1) the observation window per night (airmass $<2$)
shrinks, beginning about 15 days after opposition; 2) the asteroid
retrograde motion slows, so that it remains in the high-cadence fields
for more nights; 3) the same slowing proper motion decreases
the trailing and its consequent error degradation; 4) the asteroid
phase angle increases, thus making it fainter.  In addition, the
distance to the asteroid changes, but this effect is relatively
minor. These effects each impact the astrometric and photometric
precision differently.  
Therefore, we examine these two impacts separately.

As discussed in Section~\ref{sec:asteroids}, the astrometric precision
of combined parallax measurements scales as $(v_\rel/t_\obs)^{3/2}$.
Since both factors decline during the approach to prograde motion, they
tend to cancel.  However, in the approximation of co-planer orbits,
the first term actually goes to zero, 
which is both unphysical for extended observations
and not actually true for non-co-planar orbits.  To suppress this singularity
we adopt orbital inclinations of $10^\circ$, which is typical for
bulge ecliptic latitudes.  We then find that the ratio $(v_\rel/t_\obs)$
monotonically declines (improving parallax precision) from opposition
to the prograde boundary.  For example, it falls by a factor $\sim 1.6$
at 40 days from the Solstice, 
implying a factor 2 improvement in astrometric precision.
There is also an improvement from decreased trailing, but this
improvement is modest because the degradation at opposition is itself
minor.

However, these improvements are countered by declining brightness at 
non-zero phase angle $\alpha$.  To calculate this, we assume a flux factor
\begin{equation}
Z = (1-g)\exp[-A_1(\tan(\alpha/2))^{B_1}]
+ g\exp[-A_2(\tan(\alpha/2))^{B_2}]
\label{eqn:fluxfac}
\end{equation}
where $A_1 = 3.33$, $A_2 = 1.87$, $B_1 = 0.63$, $B_2 = 1.22$, $g = 0.15$
\citep{dymock07}. 
We then find that the net impact of all three factors on astrometric precision
is complex.  There
is overall degradation from 0 to 35 days after opposition, which peaks
at 25\% about halfway through.  Then there is increasing improvement
toward the boundary of the retrograde season.

The photometric precision scales inversely as the square root of the
exposure time, $\sim (v_\rel/t_\orb)^{1/2}$, i.e., as the cube-root of the
astrometric impact.  Curiously, as we show in the Appendix, the
impact of trailing on photometric precision also scales as the cube
root of its impact on astrometric precision in the range of interest.
However, the non-zero phase impacts astrometry and photometry 
equally.  Thus, the overall adverse impact is worse,  rising 
toward 0.45 mag at 35 days and then declining toward zero at the
season boundary.

These results imply that the magnitude limit for defining orbits
(from microlensing data alone) will vary as a  function of phase,
up or down by about 0.3 mag relative the special case of opposition,
calculated above.  Moreover, the rotation-period 
and variability errors will on average be somewhat larger than those
calculated in Section \ref{sec:rotpers}.

Finally we consider how changing the asteroid orbit impacts these
precisions.  We find that for other orbital radii, the variation
with time from opposition is qualitatively similar to the case
examined above for $x=3$, so that the main differences are found
by comparing the cases at  opposition.  Here there are two effects,
both due to the higher proper motion for closer asteroids.
First, the time spent in the high-cadence zone is reduced and 
second, the image trailing is more severe.  Together,
these nominally lead to a factor 2 decrease in astrometric precision at $x=2$
relative to $x=3$.
However, because the asteroid is closer, less astrometric precision
is actually required to determine its orbit.  The most critical
measurement is the radial velocity $v_r$, whose error scales as
$r^2$.  Hence, the orbit determination substantially improves at $x=2$.

There is also a photometric effect, but this is much smaller, roughly
a 0.25 mag degradation at $x=2$.  However, it is important to keep in mind that
asteroids of fixed physical characteristics are 9 times brighter
at $x=2$ than $x=3$.

In summary, the sensitivity of microlensing survey data to 
asteroids does vary with orbital phase and orbital radius.
However, the variations are relatively modest within the 
``retrograde season'' for main belt asteroids.  And the key point
remains: this sensitivity, while variable, is rigorously
calculable for asteroids with any definite set of properties that
one might consider.

\section{{Asteroid Recovery}
\label{sec:recovery}}

\subsection{{Orbital-period timescales}
\label{sec:orbpertime}}

The error in the orbital period
given by Equation~(\ref{eqn:perioderror})
would be small enough to characterize
the asteroid orbit for many purposes.  But it would also mean that the
asteroid could easily be recovered if it passed through the bulge
fields during microlensing season on a previous or subsequent orbit.
That is, at any given instant, there would be an essentially one-dimensional,
roughly $0.028*360^\circ\sim 10^\circ$ 
track defining the locus of possible positions for reappearance
of the asteroid.  While there might be other asteroids near this
track, the tentative identification of one such image would fix
the orbit with essentially infinite precision, which would immediately
lead to secure predictions of its position in other images.  

What fraction of asteroids could be recovered in this way?
Recall from Section~\ref{sec:frame} that about 2.8\% of all 
previously discovered asteroids
that pass through high-cadence bulge fields will be detected each
year.  Ignoring for the moment that the asteroids have a range of
inclinations, this means that an average of 2.8\% will be recovered
each year.  Of course, for asteroids on $5\pm 0.5$ year orbits, none
will be recovered during the first four years, and 14\% during the
fifth year.  Now, if an asteroid has an integer-year orbital period, then
it would return to exactly the same place in 
the high-cadence microlensing field, regardless of inclination.
However, if it is highly inclined and does not have an integer-year
period, it will likely miss the high-cadence fields.  But here it
is important to point out that in contrast to the observations 
required for high-precision orbit determination, the recovery
observations can be quite sparse.  The regions of such sparse observations
are much larger than the high-cadence regions.  Moreover, such sparse
observations over roughly 50 deg$^2$ were carried out from 2001-2009
by OGLE-III.  Hence, it is likely that tens of percent of asteroids
could be recovered using a combination of OGLE-III and OGLE-IV data.

It is quite possible that the great majority of asteroids whose
orbits can be characterized by microlensing-survey data will 
already be known and can simply be matched to entries in the
Minor Planet Center database.  For example, \citet{polishook12} carried out a
blind survey using the Palomar Transit Factory, which covered a similar
area to the high-cadence OGLE fields with 10 observations per night, for four
nights.  Of the 30
asteroids that they discovered with $18<R<18.5$ (whose faint limit 
corresponds to $I\sim 18.1$), all were already known.
In the next bin $18.5<R<19$, only 2 out of 50 were previously unknown.

\subsection{{Single-night predictions}
\label{sec:pred}}

For completeness, we estimate how well the reappearance of an asteroid
can be predicted given a single night of uniformly-spaced data, with
$n$ observations over $t_\obs$ centered at transit, 
each with $\sigma_0$ precision.  For
this purpose, we approximate the projected motion of the observatory
(relative to the Earth center)
as $R_\oplus\cos\delta\sin(\omega t)
\sim R_\oplus\cos\delta[\omega t - (\omega t)^3/6]$,
where $\omega\equiv 2\pi\,{\rm day}^{-1}$.  Then, if one
naively ignores parallax, one will make an error in the
geocentric angular velocity $\Delta \mu \simeq \omega R_\oplus\cos\delta/r$.
This implies that 
if the distance $r$ is estimated with precision $\sigma(r)$,
then the error in $\mu$ will be
\begin{equation}
\sigma(\mu) \simeq \omega R_\oplus\cos\delta{\sigma(r)\over r^2}.
\label{eqn:sigmamu}
\end{equation}
Then, after some algebra, one finds
\begin{equation}
{\sigma(r)\over r^2} = \sqrt{100800\over n}(\omega t_\obs)^{-2}
{\sigma_0\over R_\oplus\cos\delta}.
\label{eqn:sigmar}
\end{equation}
Together, these imply that the error in the predicted position
will be
\begin{equation}
\sigma(\theta) ={\rm day}\times\sigma(\mu) = 
\sqrt{100800\over n}{{\rm day}\over \omega t_\obs^2}\sigma_0=
2.8''
\biggl({n\over 10 }\biggr)^{-1/2}
\biggl({t_\obs\over 10\,{\rm hr} }\biggr)^{-2}
{\sigma_0\over 30\,{\rm mas}},
\label{eqn:distind}
\end{equation}
independent of distance $r$.

\section{{Other Surveys}
\label{sec:other}}

We have carried out our estimates using the characteristics of the
OGLE-IV survey (supplemented by OGLE-III for asteroid recovery).
There are two other current microlensing surveys, MOA and Wise,
and one planned survey that is well on its way toward commissioning,
KMTNet.  Because the MOA survey operates from New Zealand, it faces
substantially more difficult observing conditions than the OGLE survey
does in Chile.  The seeing is typically worse and the number of
nights with good transparency during the whole night fewer.
Combined, these two effects probably degrade the distance measurements
by a factor of a few. Nevertheless, the current incarnation of MOA,
with 2.2 deg$^2$ field and 1.8m telescope has been in operation since
2007.  It covers 22 fields (almost 50 deg$^2$) every 50 minutes,
which means that the asteroids remain in the MOA fields almost twice
as long as in the OGLE fields.  Hence many asteroids could be recovered.

The Wise survey \citep{wise} is carried out from Israel using a 1.0m telescope
and 1 deg$^2$ camera.  Because it is in the northern hemisphere,
its observing window is too short for effective nightly distance
measurements.  However, being in the northern hemisphere, Wise
could make crucial observations that would complement a bulge
asteroid survey carried out by either OGLE or MOA.  Wise could observe
a broad swath toward the Galactic anti-center at low cadence during
the northern winter.  To recover asteroids, it would only be necessary
to make two 2-min observations on each of two successive nights.
Hence a campaign of 3 months could cover about 1000 deg$^2$, leading
to recovery of a large fraction of asteroids over 5 years.

The KMTNet survey will be carried out by three 1.6m telescopes 
(in Chile, South Africa, and Australia) each with $2^\circ\times 2^\circ$ 
cameras.  The survey plans to monitor four fields with a 10-minute
cycle time.  Hence, near the Summer Solstice it will observe each field
roughly 120 times per night from the two good-seeing sites
(Chile and South Africa).  It could therefore plausibly reach
$2.5\log(\sqrt{120/10})= 1.4$ mag deeper than the $I=18.1$ limit estimated
above.  And, at fixed magnitude, KMTNet could measure rotation periods
much better, if this proves to be a problem for lower-cadence observations.
On the other hand fewer asteroids could be recovered from
KMTNet data alone, because it does not plan to survey outside the
high-cadence fields and because it lacks a long time baseline.

\section{{Application to Bulge Science}
\label{sec:bulge}}

Precise astrometry of asteroids passing though the bulge would
enable wide-angle astrometry of the bulge at faint magnitudes.
For example, it would permit determination of the proper-motion
gradient across the Galactic bar, an important Galactic parameter
whose measurement has never even been attempted.

\citet{sumi04} measured the proper motions of 5 million Galactic
bulge stars with $11<I<18$ over 11~deg$^2$ using OGLE-II data, and
R.\ Poleski (2013, private communication) is in the process of
measuring proper motions over about 50~deg$^2$ using OGLE-III data.
These catalogs enable a wide range of bulge science.  However, they
are intrinsically narrow-angle, i.e., tied to a local
reference frame, and so do not permit wide-angle astrometric measurements.

For example, if the bar is inclined to our line of sight by an angle
$\alpha$ and is rotating as a solid body at frequency $\Omega$, then
after some algebra one finds that the bar proper motion $\Delta\mu(\ell)$
at Galactic longitude $\ell$ (relative to $\ell=0$) is given by
\begin{equation}
\Delta\mu(\ell) = (\Omega + \mu_{\rm SgrA*})
\sin\ell(\cot\alpha\cos\ell - \sin\ell)
-\Omega_U\sin\ell(\cos\ell + \cot\alpha\sin\ell)
\label{eqn:barrot}
\end{equation}
where $\mu_{\rm SgrA*}=-6.38\,\masyr$ \citep{reid04}
is the observed proper motion of
SgrA* in the direction of Galactic rotation and 
$\Omega_U\equiv U_\odot/R_0 \simeq (10\,\kms)/(8\,\kpc)= 0.26\,\masyr$
reflects the radial motion of the Sun $U_\odot$ toward the Galactic center.
Hence, ignoring small
terms, the gradient of the proper motion is 
$d\mu/d\ell=(\Omega + \mu_{\rm SgrA*})\cot\alpha-\Omega_U$. It would be interesting
to measure both this parameter combination and any deviations from
this predicted simple behavior.

There are several possible paths to transforming narrow-angle 
into wide-angle catalogs.  One would be simply to wait for GAIA.
A second would be to tie the OGLE proper motions to 
Tycho-II \citep{hog00}.
This would require a special set of short exposures taken over
several years because essentially all Tycho-II stars are saturated
in standard OGLE data.  The precision of this approach would be 
fundamentally limited by the relatively small number of Tycho-II stars
($\sim 200\,$deg$^{-2}$), which would yield a precision of $0.20\,\masyr$
per deg$^{2}$ patch (or  $0.22\,\masyr$ if $V<10$ stars were eliminated
due to saturation even in shortened exposures).  Third, one could tie local 
OGLE astrometry to UCAC-4 \citep{zach13},
which in turn is tied to Tycho-II.  It is
not completely clear that UCAC reductions are effective in crowded
bulge fields.  Establishing this would require detailed tests.
Moreover, this procedure would face the same fundamental limitations
as the underlying Tycho-II catalog.

Asteroids with well-defined orbits provide an alternate path to
wide-angle proper motions.  For example, any asteroid found in OGLE-IV
data and recovered in OGLE-III would have an extremely well measured
period, as well as other orbital elements.  Its predicted position
relative to an ensemble of nearby clump giants from the previous orbit
would therefore be known to high accuracy for each observation, based
on OGLE-IV observations plus the period determined from matching to
one position of OGLE-III data.  The displacement of the measured position
relative to the same ensemble of clump giants in OGLE-III data would
then give the relative angular displacement over the elapsed orbital
period.

Scaling from the \citet{polishook12} detection rate, and assuming
5 years of OGLE-IV searches, a 30\% recovery rate from OGLE-III
data, and roughly 25 observations for each within OGLE-III, this
would imply roughly 11000 measurements with typical precision
of 20 mas, spread out over 50~deg$^2$.  Since the typical elapsed
time is 5 years, this leads to proper motion precision of about
$0.26\,\masyr$ per deg$^2$, which is comparable to a calibration
based on Tycho-II.


\section{{Conclusions}
\label{sec:conclude}}

We have shown that existing microlensing surveys, particularly the OGLE
survey contain a vast wealth of asteroid data from an almost completely
untapped region of parameter space.  These data could be exploited to
construct a catalog of asteroids with well-measured orbital parameters,
precise rotation periods and amplitudes, and essentially perfectly
known completeness down to $I\sim 18.1$.  Currently, there are about
 a dozen such asteroids within a magnitude of this limit, and
these do not have well-defined completeness properties.
These same data can also be used
to find asteroid binaries and find asteroid shapes.

\acknowledgments

This work was supported by NSF grant AST 1103471.
J.C. Yee is supported by a Distinguished University Fellowship
from The Ohio State University.  We thank Scott Gaudi for
insightful discussions.  We thank David Polishook,
Radek Poleski, and David Bennett
for important comments on the manuscript.  The referee, Petr Pravec,
identified a number of key issues whose investigation greatly
improved the analysis.

\appendix

\section{{Astrometry of Trailed Images}
\label{sec:streak}}

For a general point spread function (PSF), $F(x,y)$ that is normalized
to unity, and assuming photon-limited noise, the astrometric error 
$\sigma_\hatn$ in the $\hatn$ directions is given by (e.g., \citealt{gould95})
\begin{equation}
\sigma_\hatn^{-2} = \sum_{ij}{(N\Delta F_{ij} \Delta x\Delta y)^2\over
a^2(N F_{ij} + B)\Delta x \Delta y}
\label{eqn:discrete}
\end{equation}
where $N$ is the total number of source photons, $(\Delta x,\Delta y)$
is the pixel size, $B$ is the background photon surface density,
$F_{ij}$ is the mean of $F$ in pixel $(i,j)$, and $\Delta F_{ij}$
is the change in this number under the impact of a displacement
$a\hatn$.  If the pixel scale is small compared to the structure of
the PSF, then this can be evaluated using,
\begin{equation}
\sigma_\hatn^{-2} = N\int dx\int dy {(\hatn \cdot \nabla F)^2\over
F(x,y) + B/N}.
\label{eqn:cont_gen}
\end{equation}
We now consider a Gaussian PSF of width $\sigma$, that has been
trailed by total length $\epsilon$.  Since the pixel size is small,
we can without loss of generality assume that the trail is in the $x$
direction:
\begin{equation}
F(x,y) = 
{\exp(-y^2/2\sigma^2)\int_{-\epsilon/2}^{\epsilon/2} d\ell \exp(-(x+\ell)^2/2\sigma^2)
\over 2\pi \sigma^2\epsilon}
\label{eqn:fstreak}
\end{equation}
from which we derive
\begin{equation}
{dF\over d x} = 
{\exp(-y^2/2\sigma^2)\over 2\pi \sigma^4\epsilon}
\biggl[e^{-(x+\epsilon/2)^2/2\sigma^2}
-e^{-(x-\epsilon/2)^2/2\sigma^2}\biggr],
\label{eqn:dfdxstreak}
\end{equation}
and
\begin{equation}
{dF\over d y} = {-y\over\sigma^2}F(x,y).
\label{eqn:dfdystreak}
\end{equation}

We now restrict to the background-limited case, i.e.,
$F(x,y)+B/N\rightarrow B/N$ in the denominator of 
Equation~(\ref{eqn:cont_gen}).  We then obtain for the trail direction,
\begin{equation}
\sigma_{{\hat x}}^{-2} = {N^2\over 8\pi\sigma^4 B}G_x(\eta);
\qquad G_x(\eta)\equiv {1-\exp(-\eta^2)\over \eta^2};
\qquad \eta\equiv {\epsilon\over 2\sigma}
\label{eqn:streakdir}
\end{equation}
We recognize the first factor as the standard expression for the
astrometric error of a Gaussian PSF below sky.  The second term
quantifies the degradation due to trailing.  In the limit
of $\eta\ll 1$, $G_x(\eta)\rightarrow 1-\eta^2/2$, implying that
the error is fractionally degraded by $\eta^2/4$.  For the intermediate
case, $\eta=1$ (e.g., $\epsilon=1^{\prime\prime}$ and 
FWHM $\simeq 1.2^{\prime\prime}$) the
error degradation is $G_x^{1/2}\rightarrow 0.80$.  For long trails,
$G_x^{1/2}\rightarrow \eta^{-1}$.

For the non-trailed direction, the astrometric precision is
\begin{equation}
\sigma_{{\hat y}}^{-2} = {N^2\over 8\pi\sigma^4 B}G_y(\eta);
\qquad
G_y(\eta) = {\sqrt{2\pi}\over 8\eta^2}\int_{-\infty}^{\infty} dz
[\erf(z+\eta/\sqrt{2})-\erf(z-\eta/\sqrt{2})]^2
\label{eqn:nostreakdir}
\end{equation}
This cannot be evaluated in closed form, but in the three regimes
evaluated above, $G_y^{1/2}(\eta) = 1 - \eta^2/12$ ($\eta\ll 1$);
$G_y^{1/2}(1) = 0.93$; $G_y^{1/2}(\eta) = \pi^{1/4}\eta^{-1/2}$
($\eta\ll 1$).

Finally, we note that it is straightforward to show that
the photometry precision of a trailed image is degraded
by exactly the same factor $G_y^{1/2}$
as the non-trail direction astrometry.

\begin{equation}
\label{eqn:}
\end{equation}


\end{document}